\documentstyle[12pt,epsf]{article}

\def\GeV{{\;\rm GeV}}

\def\TeV{{\;\rm TeV}}
\def\etal{{\it et al.}}

\def\pl#1#2#3{
        {\it Phys.\ Lett.\ }{\bf #1}, #2 (19#3)}
\def\zp#1#2#3{
        {\it Zeit.\ Phys.\ }{\bf #1}, #2 (19#3)}
\def\pr#1#2#3{
        {\it Phys.\ Rev.\ }{\bf #1}, #2 (19#3)}
\def\np#1#2#3{
        {\it Nucl.\ Phys.\ }{\bf #1}, #2 (19#3)}
\def\ib#1#2#3{
        {\it ibid.\ }{\bf #1}, #2 (19#3)}
\def\cpc#1#2#3{
        {\it Comp.\ Phys.\ Comm.\ } {\bf #1}, #2 (19#3)}

\begin{document}

\begin{flushright}
\mbox{
\begin{tabular}{l}
    FERMILAB-PUB-98/341-T
\end{tabular}}
\end{flushright}
\vskip 1.5cm
\begin{center}
\Large
{\bf Strong radiative corrections to $W b \bar{b}$ production \\
in $p \bar{p}$ collisions} \\
\vskip 0.7cm
\large
R.K. Ellis and Sini\v{s}a Veseli \\
\vskip 0.1cm
{\small Theory Group, Fermi National Accelerator Laboratory,
P.O. Box 500, Batavia, IL 60510} \\
\vskip 0.1cm
\end{center}
\thispagestyle{empty}
\vskip 0.7cm

\begin{abstract}
We calculate the strong radiative corrections to the
process $p\bar{p}\rightarrow W(\rightarrow e\nu)g^*(\rightarrow b\bar{b})$.
At the Tevatron this process is the
largest background to the associated Higgs boson production
$p\bar{p}\rightarrow W(\rightarrow e\nu)H(\rightarrow b\bar{b})$.
The calculation is based on the subtraction procedure,
and the corrections
are found to be significant.
\end{abstract}
\newpage

\section{Introduction}

The search for the origin of electroweak symmetry breaking is at the
top of the agenda for the next generation of collider experiments. In
the standard model (SM) the mechanism of electroweak symmetry breaking
predicts the existence of a single uncharged Higgs boson whose mass is
{\it a priori} unknown. It is currently
constrained to be $90< M_H < 280\GeV$ at 95\% confidence level
\cite{ICHEP98},
where the upper bound comes from precision electroweak measurements,
while the lower bound is determined by direct
searches.\footnote{We note here that
in the minimal supersymmetric extension of the standard
model (MSSM) the Higgs boson mass  is constrained to be less
than about $130\GeV$ \cite{MH_MSSM}, and that for some range of parameters
the neutral Higgs boson of the MSSM behaves like the
Higgs boson of the standard model \cite{MSSM}.}

At the Tevatron collider there is the potential to look
for the SM Higgs boson using the decay mode
$H \rightarrow b\bar{b}$ \cite{SMW}. In the lowest order (LO) the
most promising process is
\begin{equation}
p \bar{p} \rightarrow W(\rightarrow e\nu) H(\rightarrow b \bar{b})\ .
\label{wh}
\end{equation}
This search can cover
the mass range up to about $130\GeV$,
once event samples, perhaps as large $30\mbox{fb}^{-1}$, have
been accumulated.
Collection of data samples of this size
will not be easy. However, the Tevatron search is of great
importance, especially because
the mass range between $100<M_H<130\GeV$ is one of the most challenging
regions for the LHC to look for the SM Higgs \cite{ATLAS}.

In this letter we calculate the strong radiative corrections
to the $Wb\bar{b}$ process
\begin{equation}
p \bar{p} \rightarrow W(\rightarrow e\nu) g^*(\rightarrow b \bar{b})\ ,
\label{wbb}
\end{equation}
which is the principal background for the
associated Higgs production (\ref{wh})
at the Tevatron.
Other non-negligible backgrounds, provided
by the production  of $WZ,\ t \bar{b},\ t \bar{t}$ and other processes
\cite{SMW}, will not be considered here. The calculation is
performed in the limit where $b$-quark is massless, and our results
indicate that the next-to-leading order (NLO) corrections to the $Wb\bar{b}$
background are significant.

In Section \ref{calc} we briefly describe the calculation method
based on the subtraction procedure \cite{ERT},
as formulated in \cite{CS}.
Results for the NLO corrections to the lowest order
processes (\ref{wh}) and (\ref{wbb}) at the Tevatron
are presented in Section \ref{res}, while conclusions are given in
Section \ref{conc}.

\section{Calculation method}
\label{calc}

In order to evaluate the strong radiative corrections to processes
(\ref{wh}) and (\ref{wbb})
we have to consider Feynman diagrams describing real radiation,
and also the ones involving virtual corrections to
the tree level graphs.

The corrections due to real radiation are dealt
with using the general subtraction algorithm
formulated by Catani and Seymour \cite{CS}, which is based
on the fact that the singular parts of the QCD matrix elements
for real emission can be singled out in a process-independent manner.
By exploiting this observation one can construct a set of
counter-terms that cancel
all non-integrable singularities appearing in real matrix elements. The
NLO phase space integration can then be performed numerically in four
dimensions.

The counter-terms that were subtracted from the real matrix elements
have to be added back and integrated analytically in $n$
dimensions over the phase space of the extra emitted parton,
leading to poles in $\epsilon=(n-4)/2$.
After combining those poles with the ones coming from the virtual
graphs all divergences cancel, so that one can safely perform the limit
$\epsilon \rightarrow 0$ and carry out the remaining phase space
integration numerically.

For the signal process (\ref{wh}) we consider only
the effects of the initial state gluon emission. The final state
radiation can be taken into account in the total rate by using
the radiatively corrected branching ratio for $H\rightarrow b\bar{b}$
\cite{spira}. The virtual corrections to
(\ref{wh}) are of the Drell-Yan type and
are well known \cite{AEM}.
They are expressible as a multiple of the lowest order matrix
element squared.

For the $Wb\bar{b}$ background process (\ref{wbb})
we consider real radiation from both initial and final
state quarks. The virtual corrections to the tree level graphs
can be obtained by crossing the
one loop helicity amplitudes for the process
$e^+ e^-\rightarrow \bar{q}q\bar{Q}Q$  \cite{BDKW}.

Note that our final results are presented in the
$\overline{MS}$ renormalization and factorization scheme. However,
in intermediate steps for the $Wb\bar{b}$ process we used
the four dimensional helicity scheme of \cite{BDKW}.

For the sake of simplicity,
we performed the calculation
in the limit where $b$-quark is considered massless, and with CKM
matrix elements $V_{ub}$ and $V_{cb}$ set to zero.
In the case of $Wb\bar{b}$ background, the latter
approximation eliminates the need to take into account
scattering processes involving $b$-quarks in the initial state. Given
that, for example, $|V_{cb}/V_{ud}|^2\approx 0.002$, the effects
of setting $V_{cb}=V_{ub}=0$ are small.
Corrections for the finite $b$-quark
mass are expected to be of order
$4 m_b^2/M_{b\bar{b}}^2$, or about 1\% for $M_{b \bar{b}} \sim 100\GeV$.

In order to ensure that we have a hard sub-process, we have also
introduced a set of basic cuts,
\begin{equation}
\label{hard_process_cuts}
\begin{array}{rcl}
(p_b+p_{\bar{b}})^2 &>& 4 Q^2\ , \\
p_b^T &>& Q\ , \\
p_{\bar{b}}^T &>& Q,
\end{array}
\end{equation}
where $Q$ is a scale of the same order as the $b$-quark mass.
The first constraint imposes the correct physical threshold even
though we have set the $b$ quark mass to zero.
The constraint on the $p_T$ of
the $b$ and $\bar{b}$ quarks obviates the need for factorization subtractions
involving the lowest order process $q b\rightarrow W q^\prime b$.
In general, more stringent cuts on all three quantities
will be required for comparison with experimental data.

\section{Results}
\label{res}

All results given in this paper report on the rate obtained
for $W^+$ production in $p\bar{p}$ collisions at
$\sqrt{S}=2\TeV$. To include the contributions from
$W^-$ production and the contributions from the $W^\pm$ decay into muons, one
should multiply our results by a factor of four. Note however that
we assume perfect efficiency $\epsilon_{b}$ for detection of $b$-jets.
Achievable values
of this efficiency would decrease our results
by a factor of $\epsilon_{b}^2 \approx 0.2$ \cite{BTAG}.
We used the MRSR2
parton distribution functions
with $\alpha_S(M_Z)=0.120$ \cite{MRS},
while the scale $Q$ from
(\ref{hard_process_cuts}) was set equal to $4.62\GeV$.

\begin{figure}[p]
\vspace{12.0cm}
\includegraphics{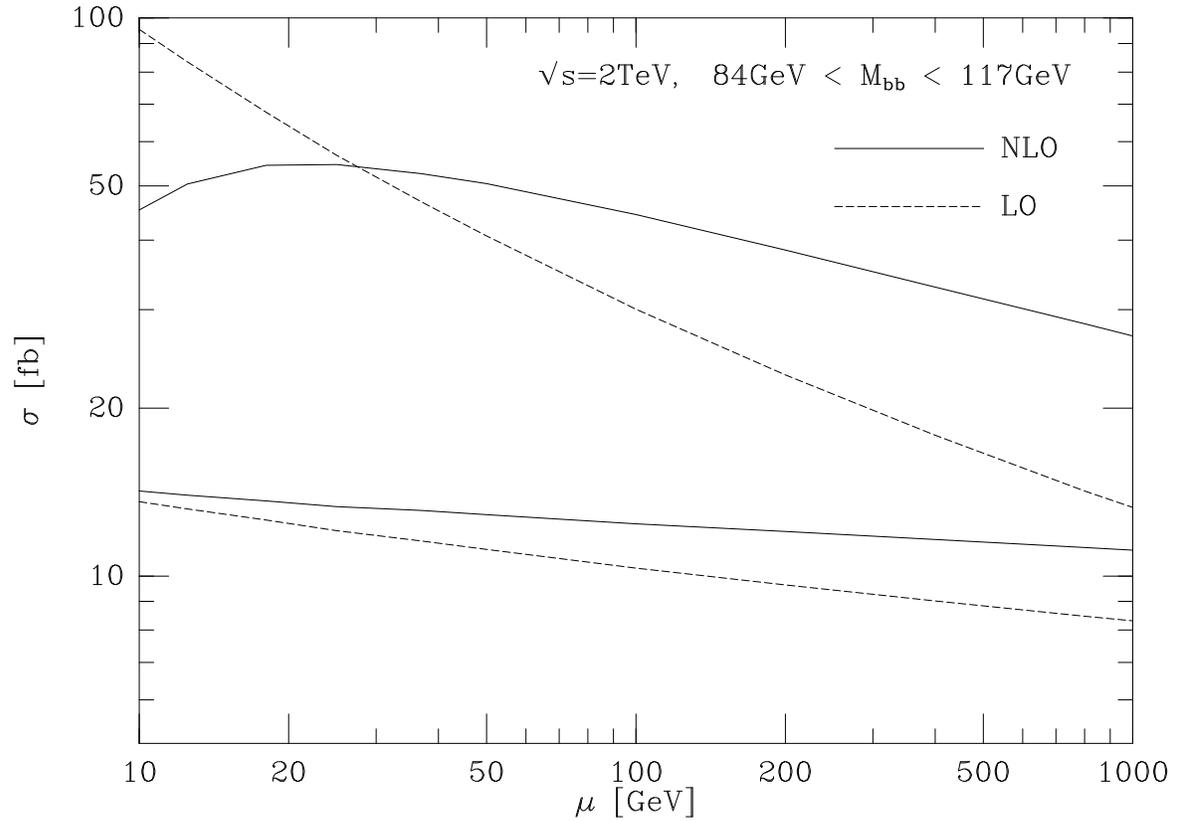}
\caption{
Cross section scale dependence for the signal (lower curves) and
$Wb\bar{b}$ background
(upper curves). These results were obtained for
$84<M_{b\bar{b}}<117\GeV$ ($M_H=100\GeV)$.
Apart from (\protect\ref{hard_process_cuts}), no other cuts have
been applied.
}
\label{mu_nocuts}
\end{figure}

In Figure \ref{mu_nocuts} we first show
the scale dependence of the LO and NLO
cross sections for the signal and background processes
in a mass window $84<M_{b\bar{b}}<117\GeV$,
which is appropriate \cite{SMW} for a $100\GeV$ Higgs boson. No
other cuts apart from (\ref{hard_process_cuts})
have been applied. At a {\em natural} scale of
$\mu=100\GeV$ we find that the K-factor is about 1.2 for the signal,
but 1.5 for the background.

As already mentioned, more strict cuts than those of
(\ref{hard_process_cuts}) have to be applied for comparison
with experiment. In addition to the cuts on rapidity
and transverse momentum,
\begin{equation}
\begin{array}{rcl}
|y_b|, |y_{\bar{b}}| &<& 2\ , \\
|y_e| &<& 2.5\ , \\
|p^T_b|, |p^T_{\bar{b}}| &>& 15\GeV \ ,\\
|p^T_e|, |p^T_{\nu}| &>& 20\GeV \ ,
\end{array}
\label{cuts1}
\end{equation}
we also impose isolation cuts,
\begin{equation}
R_{b\bar{b}},R_{eb},R_{e\bar{b}}>0.7\ ,
\label{cuts2}
\end{equation}
as well as a cut on the scattering angle of the $b \bar{b}$
system \cite{KKY} (the Higgs scattering angle)
in the Collins-Soper frame \cite{CSFrame},
\begin{equation}
|\cos{\theta_{b\bar{b}}}|< 0.8\ .
\label{cuts3}
\end{equation}
Note that imposing the cut on $\cos{\theta_{b\bar{b}}}$
requires knowledge of the
longitudinal component of a neutrino momentum. By assuming that $W$ boson is
on shell, and using $p_e$ and $p^T_{\nu}$ which are actually
measured, this component can be
reconstructed up to a two-fold ambiguity for a solution of a
quadratic equation.  Due to the asymmetry
of the neutrino rapidity distribution, by choosing the larger (smaller)
solution for $p^z_{\nu}$ in the case of $W^+$ ($W^-$),
one can improve the probability
of finding the correct $W$ momentum. Following this prescription,
in our LO (NLO) Monte Carlo studies with the above cuts
we have observed efficiency of about
$77\%$ ($68\%$) and $70\%$ ($61\%$) for processes (\ref{wh}) and
(\ref{wbb}), respectively.\footnote{The actual efficiency of
accepting an event based on the calculated value of $\cos{\theta_{b\bar{b}}}$
from the reconstructed $W$ momentum is higher. With cuts
(\protect\ref{cuts1})-(\protect\ref{cuts3}) we found that about
$93\%$ ($87\%$) and $87\%$ ($79\%$) of the LO (NLO) events
for processes (\ref{wh}) and (\ref{wbb}) were correctly accepted or rejected.}

\begin{figure}[p]
\vspace{12.0cm}
\includegraphics{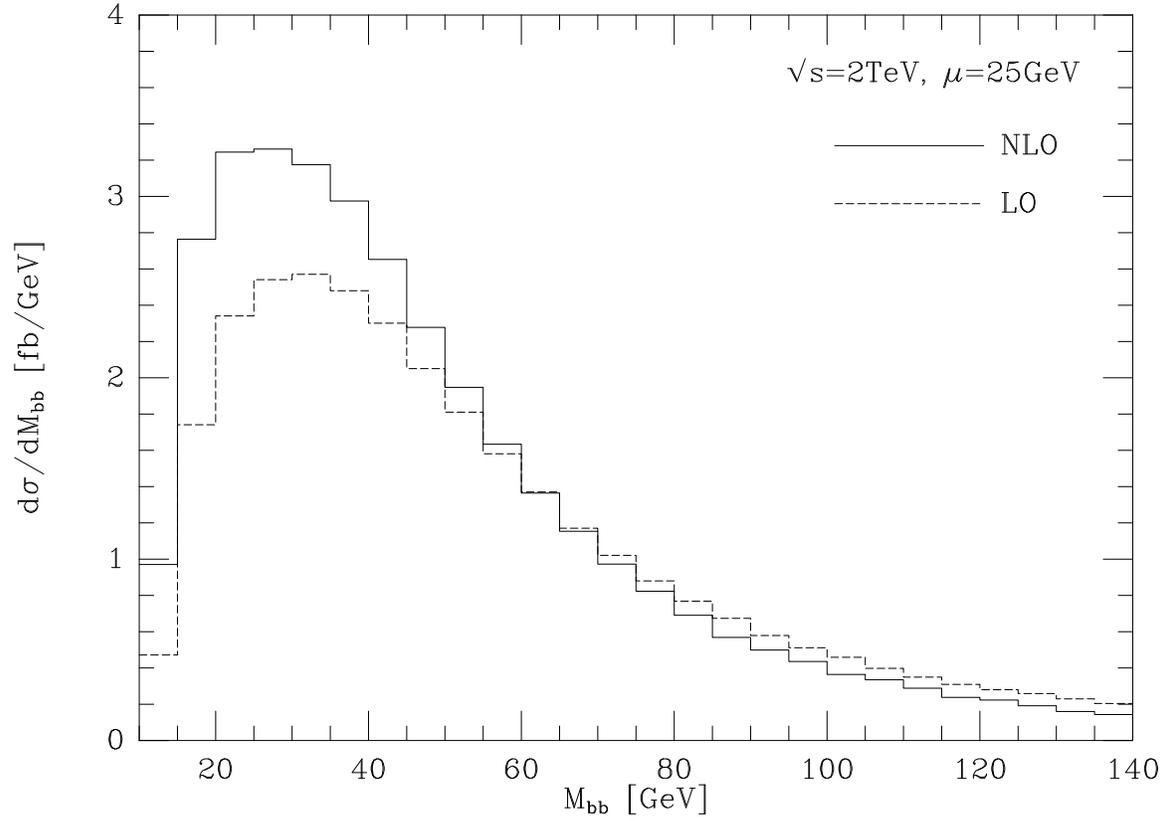}
\caption{
The LO and NLO $M_{b \bar{b}}$
dependence of the $Wb\bar{b}$ background process at a scale
of $\mu = 25\GeV$.
The results shown were obtained after imposing cuts given in
(\protect\ref{cuts1})-(\protect\ref{cuts3}).
}
\label{mbb}
\end{figure}

\begin{figure}[p]
\vspace{12.0cm}
\includegraphics{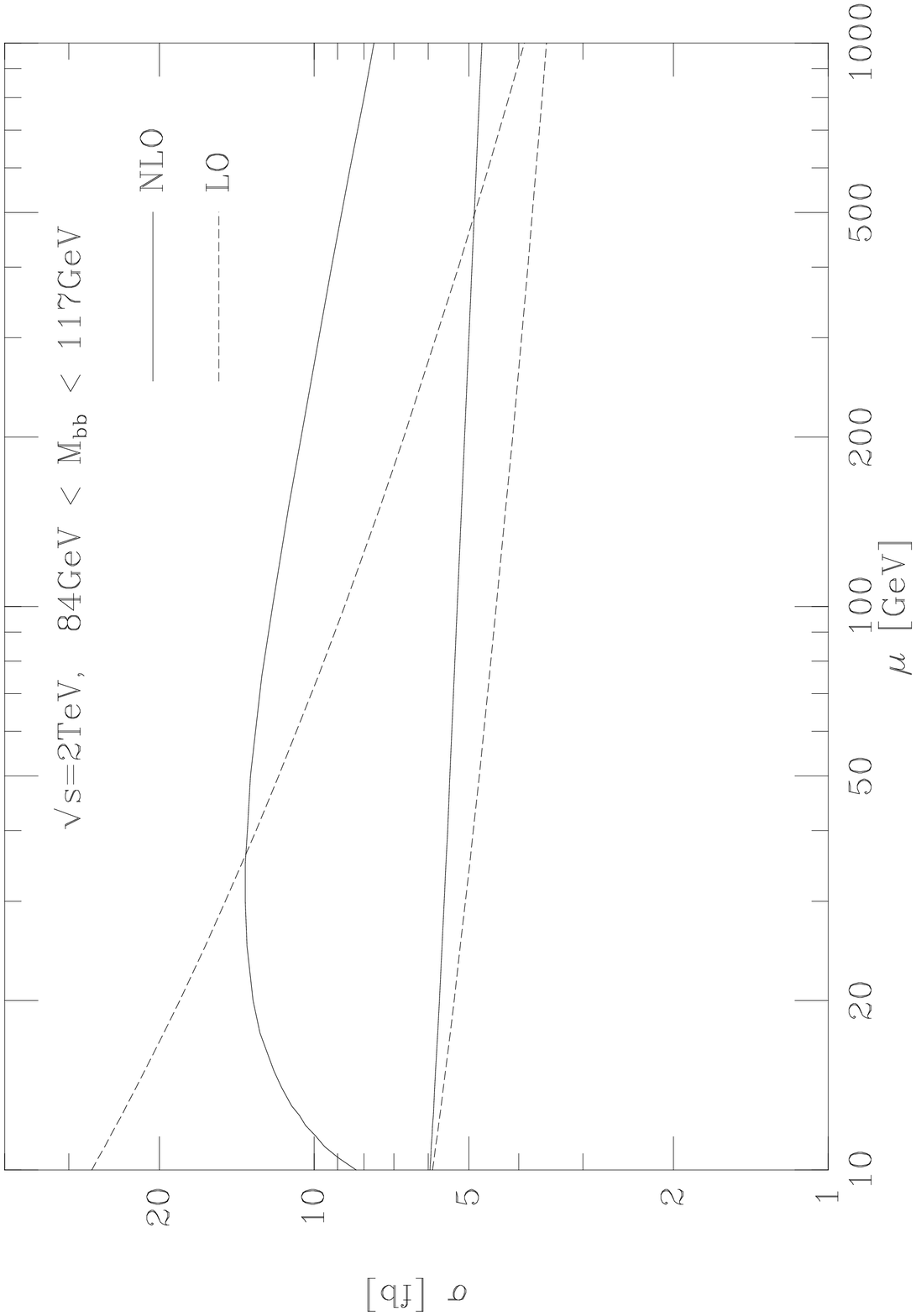}
\caption{
Cross section scale dependence for the signal (lower curves) and
$Wb\bar{b}$ background
(upper curves). These results were obtained for
$84<M_{b\bar{b}}<117\GeV$ ($M_H=100\GeV$), and after imposing cuts given in
(\protect\ref{cuts1})-(\protect\ref{cuts3}).
}
\label{mu_cuts_100}
\end{figure}

\begin{figure}[p]
\vspace{12.0cm}
\includegraphics{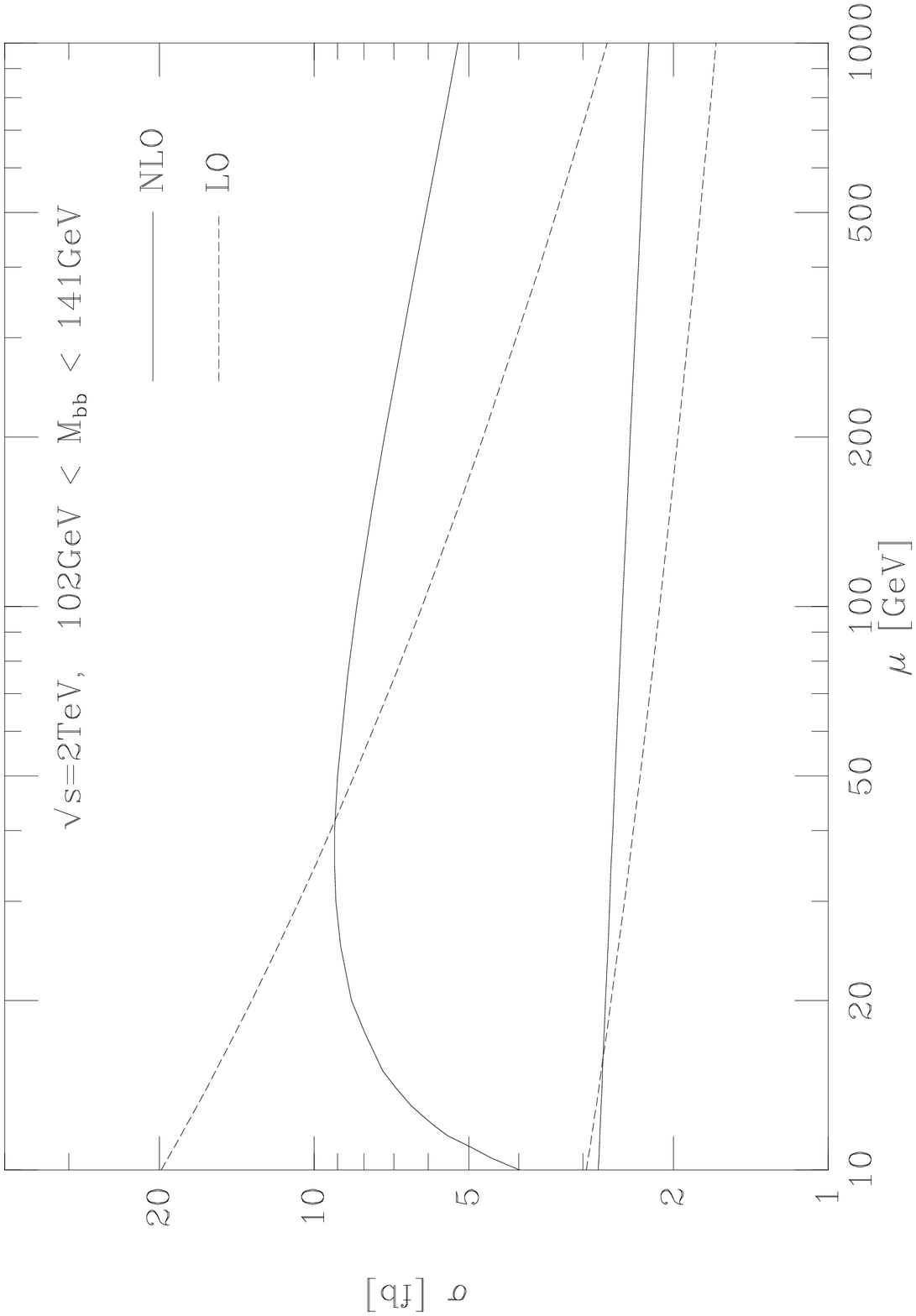}
\caption{
Cross section scale dependence for the signal (lower curves) and
$Wb\bar{b}$ background
(upper curves). These results were obtained for
$102<M_{b\bar{b}}<141\GeV$ ($M_H=120\GeV$), and after imposing cuts given in
(\protect\ref{cuts1})-(\protect\ref{cuts3}).
}
\label{mu_cuts_120}
\end{figure}

Figures \ref{mbb}, \ref{mu_cuts_100} and \ref{mu_cuts_120} show
our results obtained after cuts (\ref{cuts1})-(\ref{cuts3}) have
been imposed. In Figure \ref{mbb} we illustrate the
$b\bar{b}$ mass dependence of the $Wb\bar{b}$ process in LO and NLO.
The shape of this curve could be important in extrapolating the $W b \bar{b}$
background from observed events with lower $b \bar{b}$ invariant mass.
Note however that the we have not yet clustered the $b$-partons with emitted
radiation to form $b$-jets, although our numerical programs are
flexible enough to
allow that. This procedure could alter the shape of the $b \bar{b}$ mass
spectrum.

Figures \ref{mu_cuts_100} and \ref{mu_cuts_120} show the scale dependence
of the LO and NLO cross sections for $84<M_{b\bar{b}}<117\GeV$
($M_H=100\GeV$) and $102<M_{b\bar{b}}<141\GeV$ ($M_H=120\GeV$).
At {\em natural} scales we estimate a
signal to $Wb\bar{b}$ background ratio of 0.43 and 0.30 at
$M_H=100\GeV$ and $M_H=120\GeV$, respectively. Before definitive conclusions
can be drawn on the Tevatron search for the SM Higgs boson
other backgrounds need to be included.

\section{Conclusions}
\label{conc}

We have presented first results from a calculation of the radiative
corrections to the
production of $b \bar{b}$
in association with a $W$. The corrections are observed to be large
and positive at {\em natural} scales.
The full implications of this for the search for the standard
model Higgs boson at the Tevatron will be discussed elsewhere \cite{EV}.
Note that sizable corrections have also been found in other
reactions where the same virtual matrix elements have been used in
a crossed channel \cite{DS,NT}.
As yet we have no clear analytic understanding
of why this should be so.

Our calculation can be considered
the first step in the calculation of $W$ + 2 jet cross
section. In addition,
extension of our programs to other two boson processes
should be straightforward.

\section*{Acknowledgements}
We  are happy to acknowledge useful discussions with L. Dixon, M. Mangano,
S. Parke and M. Seymour.
Fermilab is operated by URA under DOE contract DE-AC02-76CH03000.

\end{document}